\documentclass[aps,prl,twocolumn,floats,showpacs,letterpaper,superscriptaddress,tighten]{revtex4}
\usepackage{amssymb}
\usepackage{amsbsy}
\usepackage{amsmath}
\usepackage{graphicx}
\usepackage{graphics}
\usepackage{setspace}
\usepackage[utf8]{inputenc}
\usepackage{array}
\usepackage{color}
\usepackage{fontenc}
\usepackage{textcomp}
\usepackage{rotating}
\usepackage{bm}

\let\a=\alpha   \let\d=\partial 

\let\s=\sigma

\def\bpm{\begin{pmatrix}}
\def\epm{\end{pmatrix}}
\def\be{\begin{equation}}
\def\ee{\end{equation}}
\def\bea{\begin{eqnarray}}
\def\eea{\end{eqnarray}}
\def\ba{\begin{array}}
\def\ea{\end{array}}

\def\td{\tilde}

\def\vep{{\varepsilon}}

\def\hs{\hat{\sigma}}

\def\Tr{\text{Tr}}

\def\rmd{\mathrm{d}}

\newcommand{\trasp}{\mathsf{T}}

\newcommand{\vf}{\mathsf{v}_\mathrm{F}}
\newcommand{\vc}{\mathsf{v}_\mathrm{c}}

\newcommand{\mathsym}[1]{{}}
\newcommand{\unicode}[1]{{}}

\newcommand{\bsg}{\boldsymbol{\gamma}}
\newcommand{\bss}{\boldsymbol{\sigma}}
\newcommand{\htO}{\hat{\mathcal{O}}}
\newcommand{\hsfT}{\hat{\mathsf{T}}}
\newcommand{\htH}{\hat{\mathcal{H}}}
\newcommand{\hTh}{\hat{\Theta}}

\newcommand{\TR}{\mathcal{T}}
\newcommand{\ket}[1]{\left| {#1} \right\rangle}

\newcommand{\ord}[1]{\bm{\mathit{O}}\left(#1\right)}

\begin{document}
\title{Topological Protection from Random Rashba Spin-Orbit Backscattering: \\
	Ballistic Transport in a Helical Luttinger Liquid}
\author{Hong-Yi~Xie} \email{hongyi.xie@rice.edu}
\affiliation{Department of Physics and Astronomy, Rice University, Houston, Texas 77005, USA}
\affiliation{Kavli Institute for Theoretical Physics, University of California, Santa Barbara, California 93106, USA}
\author{Heqiu Li}
\affiliation{Department of Physics, Zhejiang University, Hangzhou, Zhejiang 310027, China}
\author{Yang-Zhi~Chou}
\affiliation{Department of Physics and Astronomy, Rice University, Houston, Texas 77005, USA}
\author{Matthew~S.~Foster}
\affiliation{Department of Physics and Astronomy, Rice University, Houston, Texas 77005, USA}
\affiliation{Kavli Institute for Theoretical Physics, University of California, Santa Barbara, California 93106, USA}
\affiliation{Rice Center for Quantum Materials, Rice University, Houston, Texas 77005, USA}
\date{\today\\}

\begin{abstract}
The combination of Rashba spin-orbit coupling and potential disorder induces a random 
current operator for the edge states of a 2D topological insulator.  We prove that 
charge transport through such an edge is ballistic at any temperature, 
with or without 
Luttinger liquid interactions. The solution exploits a mapping to a spin 1/2 in a 
time-dependent field that preserves the projection along one randomly undulating component 
(integrable dynamics). Our result is exact and rules out random Rashba backscattering 
as a source of temperature-dependent transport, absent integrability-breaking terms.
\end{abstract}

\maketitle
The edge states of a quantum spin Hall (QSH) insulator realize a time-reversal 
symmetric helical Luttinger liquid (HLL): Two counterpropagating modes 
possess 
opposite spins 
and hence form a Kramers 
doublet~\cite{KaMe2005a,KaMe2005b,WBZ2006,BeZh2006,BHZ2006,QiZh2010,HaKa2010}. 
Owing to the nontrivial bulk $\mathbb{Z}_2$ topology, the HLL provides an alternative realization 
for a quantum wire with strong spin-orbit coupling and 
an odd number of channels~\cite{miev2008},
distinct from (e.g.)
carbon nanotubes~\cite{Suan2002,AnSu2002,Ta2004,SaTa2005,CaMa2006,OGM2007}.  
In an ideal QSH insulator, the $S_z$ component of the electron spin is conserved. 
Elastic backscattering of HLL edge carriers off of impurities is prohibited by 
the combination of spin U(1) and time-reversal symmetries; 
only forward-scattering potential disorder is allowed. 
The combination of pure potential disorder and Luttinger liquid interactions bosonizes 
\cite{Gogolin,ZinnJustin}
to a trivial free theory,
leading to the prediction that edge electrons exhibit ballistic transport at any temperature~\cite{BHZ2006}.  
These conclusions obtain in a \emph{fixed realization} of the disorder, a robust version
of topological protection that also applies to the surface states of 3D topological superconductors 
\cite{Ludwig08,FXC2014,XCF2015}. 

Axial spin symmetry in topological insulators is, however, \emph{not} typically 
robust~\cite{KaMe2005a,KaMe2005b,glazman2012}. 
Rashba spin-orbit coupling (RSOC) arises whenever inversion symmetry is broken, as in HgTe/CdTe \cite{Konig07} and 
InAs/GaSb \cite{Inez11,Inez14} heterostructures. 
The helical edge states then exhibit a twisted spin texture \cite{glazman2012,rachel2015}. 
Neglecting the gapped bulk, electron annihilation operators at the edge can be expanded as
\begin{align}\label{MicroMacroOp}
\begin{aligned}
	c_{\uparrow}(x) \simeq&\, e^{i k_F x} R(x) 	- i \zeta e^{-i k_F x} \partial_x L(x),
	\\
	c_{\downarrow}(x) \simeq&\, e^{-i k_F x} L(x)	- i \zeta^* e^{i k_F x} \partial_x R(x),	
\end{aligned}
\end{align}
where $R(x)$ and $L(x)$ destroy right- and left-moving edge mode electrons near the Fermi points $\pm k_F$. 
The parameter $\zeta$ encodes the strength of the RSOC. In the model of 
Refs.~\cite{glazman2012,rachel2015} $\zeta = 2 k_F / k_0^2$, where $k_0$ sets the scale for rotation 
of the spin axis. In Eq.~(\ref{MicroMacroOp}), we choose the 
quantization axis to coincide with the Kramers pair at $k = \pm k_F$. 
To lowest order in $\zeta$, the electron density operator is given by
\begin{align}\label{density}
	\rho
	=&\,
	c^\dagger_\uparrow
	c_\uparrow
	+
	c^\dagger_\downarrow
	c_\downarrow	
	\simeq
	R^\dagger R + L^\dagger L
	\nonumber\\
	&\,
	-
	\left\{
	i
	\zeta
	e^{-2 i k_F x} 
	\left[
		R^\dagger \partial_x L
		-
		(\partial_x R^\dagger) L	
	\right]
	+
	\text{H.c.}
	\right\},
\end{align}
where H.c.\ denotes the Hermitian conjugate. 
In a spinless Luttinger liquid, the time reversal operation $\TR$ exchanges $R \leftrightarrow L$ ($\TR^2 = +1$).
The term on the second line of Eq.~(\ref{density}) is odd under this, and cannot contribute to $\rho$ in the spinless case.
This term is even under time reversal in the HLL, which sends $R \rightarrow L$ and $L \rightarrow -R$ ($\TR^2 = -1$).

Two key attributes of HLLs with RSOC follow from Eq.~(\ref{density}). 
First, scalar potential disorder that couples to $\rho(x)$ generically induces a random backscattering 
component to the Dirac current operator, in the low-energy effective field theory of the HLL edge. 
Second, the screened Coulomb interaction $\rho^2(x)$ induces the usual
Luttinger liquid interaction, as well as a one-particle umklapp interaction term. 
This interaction is irrelevant in the RG sense, due to an extra derivative. 

A recent experiment \cite{Marcus15} has raised concerns that trivial edge states can 
masquerade as HLLs, while previous experiments \cite{Konig07,Inez11,Inez14,RRD15}
have not shown the anticipated ballistic transport at the lowest temperatures ($T$) for sufficiently 
long edges. A crucial theoretical task is to identify and understand mechanisms that might 
weaken topological protection and suppress the conductance at finite and zero $T$
\cite{WBZ2006,glazman2012,berkooz2012,mirlin2014,foster2015,japp2010,tra2014c}.
Although irrelevant, the one-particle umklapp interaction can be the dominant source of 
inelastic backscattering for $k_B T$ much less than the bulk gap in an isolated HLL, 
leading to $T$-dependent corrections to the edge conductance \cite{mirlin2014,foster2015,glazman2012,berkooz2012}.
Phonon scattering~\cite{tra2012b}, 
Kondo impurities~\cite{MLO2009,TFM2011,AAY2013,YWYA2015}, 
or charge puddles~\cite{VGG2013,VGGG2014} can also give $T$-dependent corrections to transport.
In this Letter we ignore these 
known mechanisms
and focus upon 
the random Dirac current operator in a disordered HLL with RSOC.

Quenched disorder that couples to the backscattering kinetic operator on the second line of 
Eq.~(\ref{density}) has been termed ``random RSOC'' in previous studies 
\cite{japp2010,tra2014c}.
Unlike backscattering (random mass) disorder in a spinless Luttinger liquid, 
short-range correlated random RSOC is irrelevant in the RG sense
for an edge Luttinger parameter $K > 1/2$ \cite{tra2014c}. 
Both the random RSOC and one-particle umklapp interaction can be simultaneously irrelevant,
and map to similar operators in bosonization \cite{japp2010,mirlin2014,SupMat}.
This suggests that both can be treated with perturbation theory, using 
bosonization to incorporate Luttinger liquid effects. Within this framework, 
random RSOC is predicted to give a $T$-dependent correction to transport that vanishes at $T = 0$
for $K > 1/2$ \cite{tra2014c}.
Moreover, it has been argued that for $K < 1/2$ the random RSOC can induce 
Anderson localization \cite{japp2010}.  
We show here that these conclusions are incorrect and miss important physics. 

In this Letter we prove that charge transport is perfectly ballistic with Landauer conductance
$G = e^2/h$ per edge, for a HLL with random RSOC at $T \geq 0$, with or without 
Luttinger interactions. We first solve the noninteracting problem 
exactly by transfer matrix, which is unitary up to a certain factor. This unitary matrix is equivalent to the 
evolution operator of a spin-1/2 magnetic moment in a random, two-component  
time-dependent magnetic field. The dynamics are integrable, since the evolution preserves
the spin projection along one randomly undulating component of the field \cite{barnes2013}, and
this translates into the absence of backscattering \cite{KaMe2005b} 
for an edge connected to ideal leads.
With Luttinger interactions, we map the problem onto one with a homogeneous 
(inhomogeneous) current operator (density-density interaction). The transformed theory is equivalent to a 
free Luttinger liquid, but with inhomogeneous Luttinger and charge velocity parameters. 
We corroborate these results with a numerical treatment of the edge wave functions and level statistics.
Finally, we compare to a disordered, particle-hole symmetric spinless quantum wire \cite{miev2008},
which also evades Anderson localization in 1D.


\emph{Model with random RSOC.}---In terms of the two-component Dirac spinor 
$\Psi(x) \equiv \begin{bmatrix} R(x) & L(x) \end{bmatrix}^\trasp$, 
the Hamiltonian of a 
noninteracting edge
incorporating random RSOC 
\cite{mirlin2014,  sh2003, goiv2004, Roth2010, japp2010, tra2012b, tra2014c}
can be written as 
\begin{subequations} \label{ham-s}
\begin{align} 
	& H_0 =   \int d x \, \Psi^\dagger \, \hat{h} \, \Psi(x),                                  				 \label{ham-1} \\
	& \hat{h} = \hat{j}(x) \left( -i \d_x \right) - {\textstyle{\frac{i}{2}}} 		\d_x \hat{j}(x) + V(x).      	 \label{ham-2}
\end{align}
\end{subequations}
In Eq.~(\ref{ham-2}) the electric current operator reads 
\be \label{def-current}
\hat{j}(x) = \bsg(x) \cdot \hat{\bss},
\ee
where 
$\hat{\bss} = \begin{pmatrix} \hs^{1}, & \hs^{2}, &  \hs^3 \end{pmatrix}$ 
are the Pauli matrices and 
$\bsg(x) = \begin{bmatrix} \xi_1(x), & \xi_2(x), & \vf(x) \end{bmatrix}$ 
encodes the random RSOC backscattering strengths $\xi_{1,2} (x)$ 
and the (possibly inhomogeneous) Fermi velocity $\vf(x)$; 
$V(x)$ denotes the forward-scattering scalar potential. 
All scattering strengths are real functions.  
Equation~(\ref{ham-s}) is invariant under time reversal $\TR$, defined by 
$\Psi(x) \to i \hs^2 \Psi(x)$ 
and 
$i \to -i$. 
The term 
in Eq.~(\ref{ham-2})
involving $\d_x \hat{j}(x)$ is required by Hermiticity
[cf.\ Eq.~(\ref{density})].


\emph{Transfer-matrix solution.}---The single-particle Schr\"{o}-dinger equation takes the form 
\be 
	\hat{h} \, \psi(x) = \vep \, \psi(x), 
\ee
where $\psi(x)$ is the two-component wave function with eigenenergy $\vep$. 
We define the current norm $\| \hat{j}(x) \|$ and the normalized current operator $\hat{J}(x) $ as
\be
	\| \hat{j}(x) \| \equiv  |\bsg(x)|, \quad
	\hat{J}(x) \equiv  \hat{j}(x) / \| \hat{j}(x) \|,
\ee
where $\| \hat{j}(x) \|$ is the local random speed. 
In terms of the rescaled wave function 
$
	\varphi (x) \equiv \sqrt{\| \hat{j}(x) \|} \, \psi(x)
$, 
the Schr\"odinger equation transforms to 
\be   \label{ham-psi}
\left(-i \d_x \right) \varphi(x)  = \htH_{\vep}(x) \, \varphi(x), 
\ee 
where $\htH_{\vep}(x)$ is composed of two components, 
\begin{subequations}  \label{def-Oe}
\begin{align}
&\, \htH_{\vep}(x) =  \htO_{\vep}(x) + \hTh(x), \\
&\, \htO_{\vep}(x) = \left[ \vep - V(x) \right] \frac{\hat{J}(x)}{\| \hat{j}(x) \|},  \quad \hTh(x) = \frac{i}{2} \hat{J} \d_x \hat{J}(x).
\end{align}
\end{subequations}
$\htO_{\vep}(x)$, $\hTh(x)$, and $\htH_{\vep}(x)$ are all Hermitian operators.
  
The transfer-matrix solution for the single-particle wave functions is
\begin{subequations} \label{Tm-solv}
\be \label{Tm-psi}
	\psi_{\vep,a}(x) 
	= 
	\left[1 / \sqrt{\| \hat{j}(x) \|} \right] \hsfT_\vep (x,-\infty) | a \rangle,
\ee
where 
$
	|a=1 \rangle \equiv \begin{pmatrix}1 & 0 \end{pmatrix}^\trasp
$ 
and 
$
	| a=2 \rangle \equiv \begin{pmatrix} 0 & 1 \end{pmatrix}^\trasp
$ 
label the degenerate Kramers pair, and the \emph{unitary} transfer matrix generated 
by the Hermitian operator (\ref{def-Oe}) reads
\be \label{Tm-form}
	\hsfT_\vep (x,x^\prime) \equiv \mathcal{P} \exp\left[ i \int_{x^\prime}^{x} \!\! \rmd{y} \, \htH_{\vep}(y) \right]. 
\ee 
\end{subequations}
Here ``$\mathcal{P}$'' denotes path ordering.
The normalization constant of the wave function (\ref{Tm-psi}) is fixed in order to recover the 
RSOC-free physics, e.g., the density profile given by the U(1) axial anomaly in (1+1) dimensions 
[see Eq.~(\ref{dens-pro})]. Using the Heisenberg equation of motion for the transfer matrix 
$
	(-i \d_{x}) \, \hsfT_\vep (x,x^\prime)= \htH_{\vep}(x) \, \hsfT_\vep (x,x^\prime)
$, 
one can prove the nontrivial relation
\be  \label{JTTJ}
	\hat{J}(x) \, \hsfT_\vep(x,\,x^\prime) = \hsfT_\vep(x,\,x^\prime)\, \hat{J}(x^\prime),
\ee 
which implies the \emph{integrability} of the transfer matrix, as discussed below.

\begin{figure}
\centering
\includegraphics[width=0.4\textwidth]{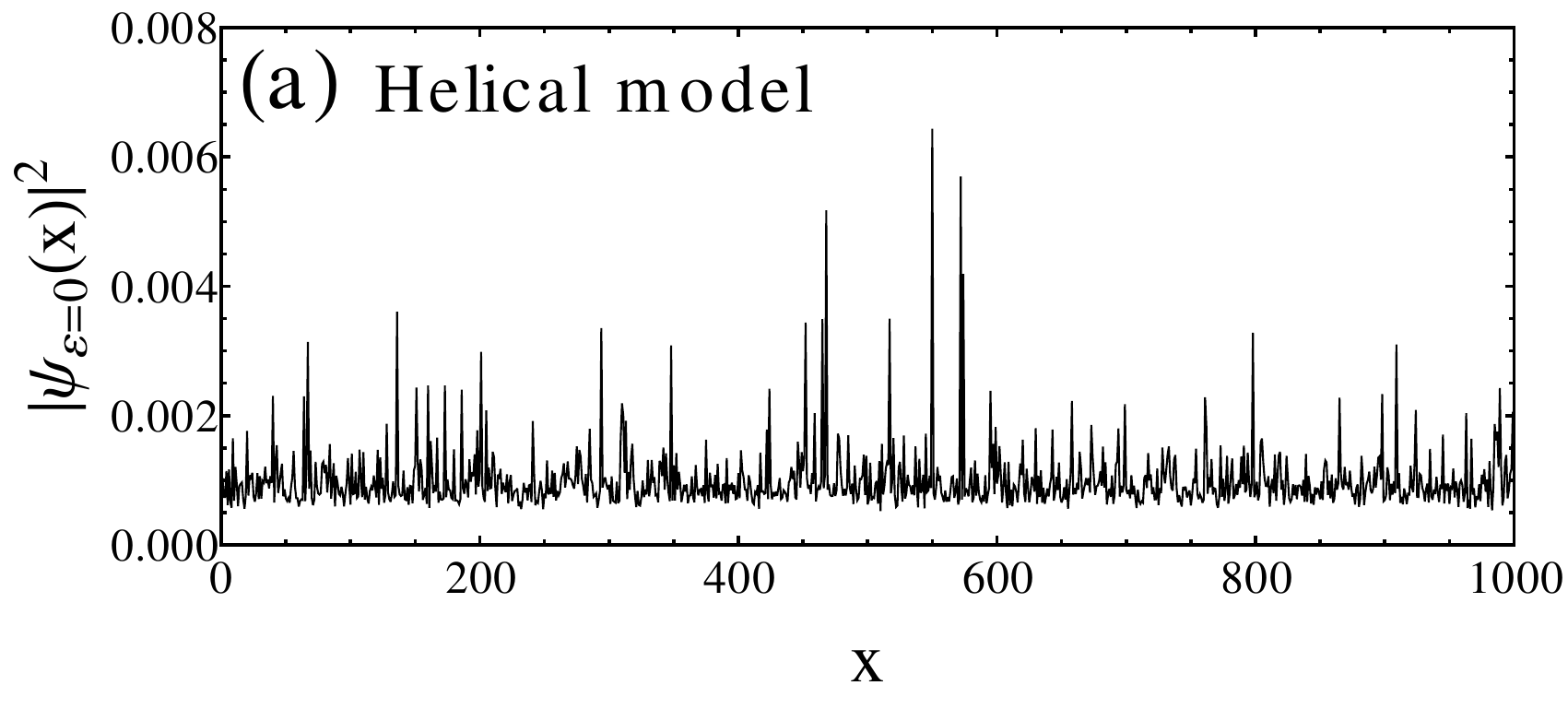}	 
\includegraphics[width=0.4\textwidth]{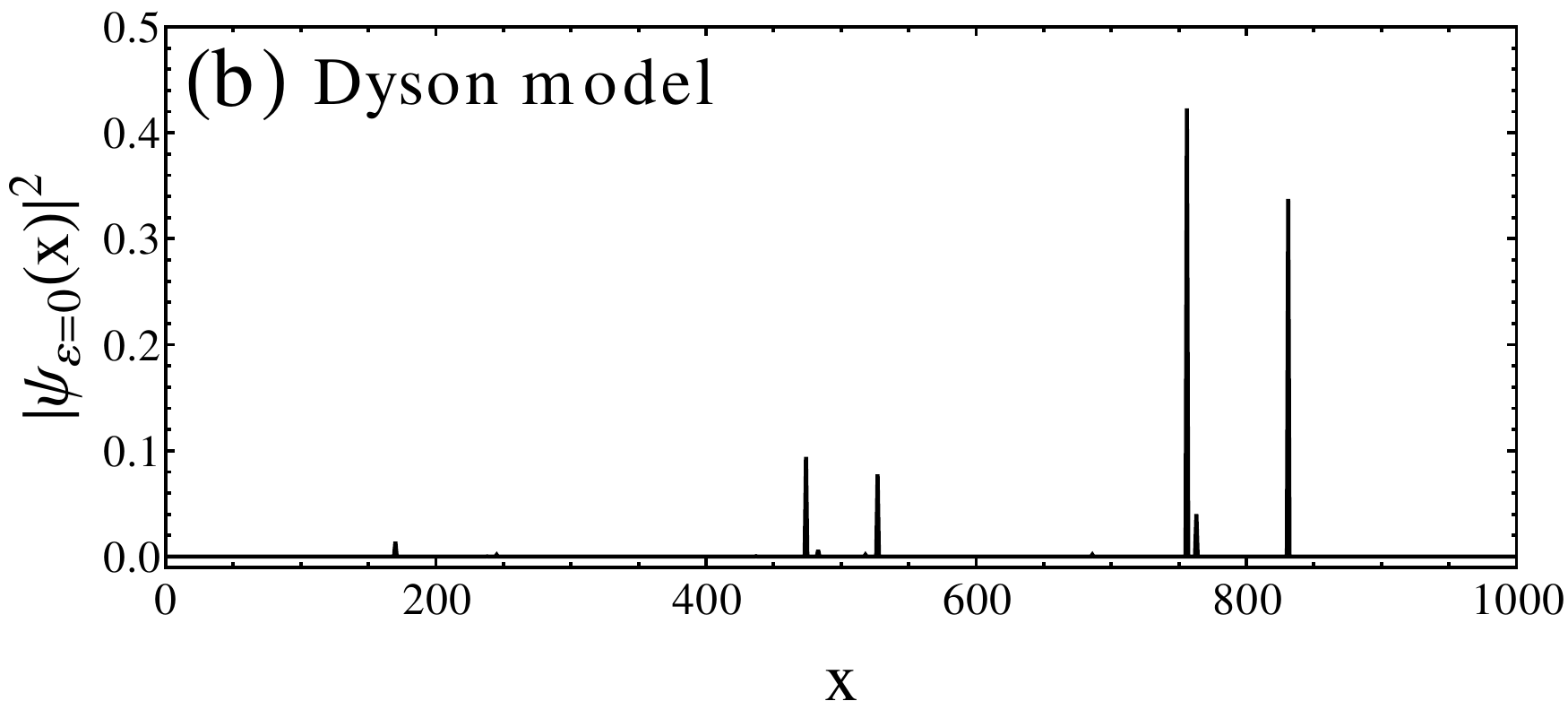}	 \\
\caption{Typical single-particle delocalized wave functions for the helical model [(a)] and the Dyson model [(b)] at energy $\vep = 0$
(momentum space exact diagonalization \cite{SupMat}).}  
\label{wfs_comparision}
\end{figure}

From the solution (\ref{Tm-solv}) we obtain 
the following conclusions:
(i) The single-particle wave functions are extended and uniformly inhomogeneous (not rarely peaked or multifractal), 
with a probability density determined by the local random 
speed 
[see also Fig.~\ref{wfs_comparision}(a)], 
\be \label{psiProfile}
	|\psi_{\vep,a}(x)|^2 \equiv \psi_{\vep,a}^\dagger (x)\, \psi_{\vep,a}(x) = \| \hat{j}(x) \|^{-1}. 
\ee 
(ii) The density profile is defined via 
$
	n(x) \equiv \lim_{\eta \to 0} \sum_a \int d \vep \, f(\vep) \, \psi_{\vep,a}^\dagger(x-\frac{\eta}{2}) \, \psi_{\vep,a}(x+\frac{\eta}{2})
$, 
where 
$
	f(\vep)=1/\left(e^{\vep/T} + 1\right)
$ 
is the Fermi-Dirac distribution. We recover the U(1) axial anomaly in (1+1) 
dimensions~\cite{Gogolin,ZinnJustin} renormalized by the local speed,
\be \label{dens-pro} 
	n(x)= - V(x) / \left[\pi  \| \hat{j}(x) \| \right]. 
\ee 
The absence of additional terms due to the random $\| \hat{j}(x) \|$  suggests that 
density-density interactions will not induce quantum (Altshuler-Aronov) corrections to transport
\cite{AA1985,XCF2015},
as we confirm below. 
(iii) The Kubo formula for the dc conductivity can be calculated via 
$	
	\s_0 = \frac{e^2}{2 h L}\int_{-\infty}^{\infty} d \vep \left[-\frac{d f(\vep)}{d \vep}\right] 
	\int d x \, d x' \, F_{\vep}(x,x^\prime)
$, 
where $L \to \infty$ is the system size and 
$
	F_{\vep}(x,x^\prime) = \Tr[\hat{J}(x) \hsfT_\vep (x,x^\prime)\hat{J}(x^\prime) \hsfT_\vep^\dagger (x,x^\prime)]
$. 
Equation~(\ref{JTTJ}) implies that $F_\vep(x,x^\prime) = 2$, independent of $x,x'$.
Then, the Kubo formula suggests a temperature-independent, universal Landauer conductance 
\be  \label{b-cond}
	G_0 = \s_0 / L = e^2 / h.
\ee
The direct calculation of $G_0$ for a HLL with random RSOC connected to ideal leads confirms this
result, as we now explain.


\emph{Integrable dynamics of a spin $1/2$ in a random, but correlated magnetic field.}---The purely 
ballistic transport in Eq.~(\ref{b-cond}) can be interpreted in terms of the instantaneous 
eigenstates of a spin $1/2$ evolving in a time-dependent magnetic field, since
the integration of the transfer matrix [Eq.~(\ref{Tm-form})]
between ideal leads is described by a corresponding spin rotation. We introduce the Hamiltonian
\begin{subequations} \label{spin-ham}
\be
	\hat{H}(t) = \sum_{\a=1}^{2}\hat{H}_\a(t), \quad \hat{H}_\a(t) = \mathbf{B}_\a(t) \cdot \hat{\bss},
\ee
with the magnetic fields $\mathbf{B}_1(t) \perp \mathbf{B}_2(t)$ defined by
\be
	\mathbf{B}_1(t) = B_1(t) \, \mathbf{n}(t), \quad \mathbf{B}_2(t) = {\textstyle{\frac{1}{2}}} \mathbf{n}(t) \times \d_t\mathbf{n}(t),
\ee
\end{subequations}
where $B_1(t)$ and $\mathbf{n}(t)$ denote the magnitude and the direction of $\mathbf{B}_1(t)$, respectively. 
The connection between the spin model~(\ref{spin-ham}) and the 
edge 
model in Eq.~(\ref{def-Oe}) becomes manifest 
if we choose $\mathbf{B}_1(t) = \frac{1}{2} \Tr[\htO_\vep(t) \hat{\bss}]$. 
The spin Hamiltonian $\hat{H}(t)$ precisely takes the form of Eq.~(\ref{def-Oe}) with time $t$ replaced by the
spatial coordinate $x$.

The time evolution of the spin is determined 
by the unitary operator
$
	\hat{U}(t) = \mathcal{T} \exp[-i \int_{0}^{t} \! \rmd{t^\prime} \, \hat{H}(t^\prime)],
$
where ``$\mathcal{T}$'' denotes time ordering. 
One can show the following for a differentiable but otherwise arbitrary field $\mathbf{B}_1(t)$: 
Starting from an eigenstate $\varphi(0)$ of $\hat{H}_1(0)$, 
$\varphi(t) = \hat{U}(t) \, \varphi(0)$ remains an instantaneous eigenstate of $\hat{H}_1(t)$. 
This statement is equivalent to the relation 
$
	\hat{U}^\dagger(t) \, \mathbf{n}(t) \cdot \hat{\bss} \, \hat{U}(t) = \mathbf{n}(0) \cdot \hat{\bss}
$, 
which is Eq.~(\ref{JTTJ}) in the spin language. 

In particular, setting 
$
	\mathbf{B}_1(0) = \mathbf{B}_1(T) = B \hat{z}
$ 
and the initial state $\varphi(0)= \ket{\uparrow}$ (aligned along $z$), 
we obtain $\varphi(T) =\varphi(0)$ for any smooth $\mathbf{B}_1(t)$ in $t \in (0,\,T)$; i.e., there is no net rotation.
Similarly, for an 
edge
with random RSOC connected to ideal leads at $x = \pm L/2$, the current operator in Eq.~(\ref{def-current}) 
satisfies $\hat{j}(\pm L/2) = \vf \, \hat{\sigma}^3$ ($\vf$ is the uniform Fermi velocity in the leads).
The transfer matrix in Eq.~(\ref{Tm-form}) is therefore reflectionless,  
and this holds for all eigenenergies $\vep$ [Eq.~(\ref{def-Oe})]. 
Since the true eigenstate $\psi$ differs from $\varphi$ only by a local Jacobian factor $\| \hat{j} \|^{1/2}$, 
identical on either side of the leads $\| \hat{j}\|^{1/2} = \sqrt{\vf}$, the transmission coefficient 
is exactly unity so that backscattering is prohibited. 
We conclude that topological protection here 
corresponds to a special, integrable two-level system that depends upon an 
arbitrary random field, and which preserves the projection along this field.
This is an explicit example of how perfect transmission is achieved for
noninteracting edges in the presence of RSOC \cite{KaMe2005b}.


\emph{Luttinger interactions.}---We consider the model in Eq.~(\ref{ham-s}) for an edge spanning $|x| \leq L/2$, connected to ideal leads.
The Hamiltonian incorporating Luttinger interactions is given by
\be\label{ham-int}
	H = H_0 + \int d x \, U(x) \, \left[ \Psi^\dagger \Psi(x) \right]^2,
\ee
where $U(x) = U \, \theta(L/2 - |x|)$, and $\theta(x)$ is the unit step function. 
Inspired by the transfer-matrix solution (\ref{Tm-psi}), we introduce the rotated fermion field 
\be  \label{tr-phi}
	\Phi(x) \equiv \sqrt{ \| \hat{j}(x) \| / \vf } \;\; \hsfT_0^{\dagger}(x,-L/2) \, \Psi(x).
\ee  
The operators $\Phi(x)$ and $ \Phi^\dagger(x^\prime)$ satisfy rescaled canonical anticommutation 
relations since the rotation in Eq.~(\ref{tr-phi}) is \emph{nonunitary}. However, one is free to perform this 
transformation in a path integral formalism, because Eq.~(\ref{tr-phi}) is a linear change of variables, 
up to a disorder-dependent Jacobian that cancels between numerator and denominator for a correlation function.
Note that $\Phi(x)  = \Psi(x)$ for $|x| > L/2$. Exploiting Eq.~(\ref{JTTJ}), Eq.~(\ref{ham-int}) reduces to 
\be  \label{ham-Phi}
	H =  \int d x \, \left\{\Phi^\dagger \vf \hat{\sigma}^3 (-i\d_x) \Phi(x) + \td{U}(x) \left[ \Phi^\dagger \Phi(x) \right]^2 \right\},
\ee
where $\td{U}(x) \equiv \vf^2 \, U(x) / \| \hat{j}(x) \|^2$. 
This transformed theory with a homogeneous (inhomogeneous) kinetic term (Luttinger interaction)
is equivalent to a free boson theory \cite{Gogolin,ZinnJustin}.  
The conductance in a Landauer setup for the $\Phi(x)$ fermions is therefore given by Eq.~(\ref{b-cond}), 
\emph{independent} of $\td{U}(x)$~\cite{MaSt1995,Pono1995,SaSch1995}. 
Bosonizing 
Eq.~(\ref{ham-Phi})
in a fixed realization of disorder gives the Luttinger parameter $K(x)$ and charge velocity $\vc(x)$ \cite{SupMat},
\be\label{LutParam}
\begin{aligned}
	K(x) =&\, 1 / \sqrt{[1 + \chi(x)][1 - \chi(x) + 2 \td{U}(x) / \pi \vf]},
	\\
	\vc(x) =&\, \vf \sqrt{[1 - \chi(x) + 2 \td{U}(x) / \pi \vf ] / [1 + \chi(x)]},
\end{aligned}
\ee
where $\chi(x) \equiv \vf / \| \hat{j}(x) \|  - 1$. 
As usual, while the mapping to a free boson parametrized by $K(x)$ and $\vc(x)$ is exact, 
we expect that the explicit formulas in Eq.~(\ref{LutParam}) are correct only to linear 
order in the perturbations $\chi$ and $\td{U}$ \cite{Gogolin}. In the noninteracting case $\td{U} = 0$,
we have $K(x) \simeq 1 + \ord{\chi}^2$ and $1 / \vc(x) \simeq 1 / \| \hat{j}(x) \| + \ord{\chi}^2$.


\begin{figure}[b]
\centering
\includegraphics[width=0.4\textwidth]{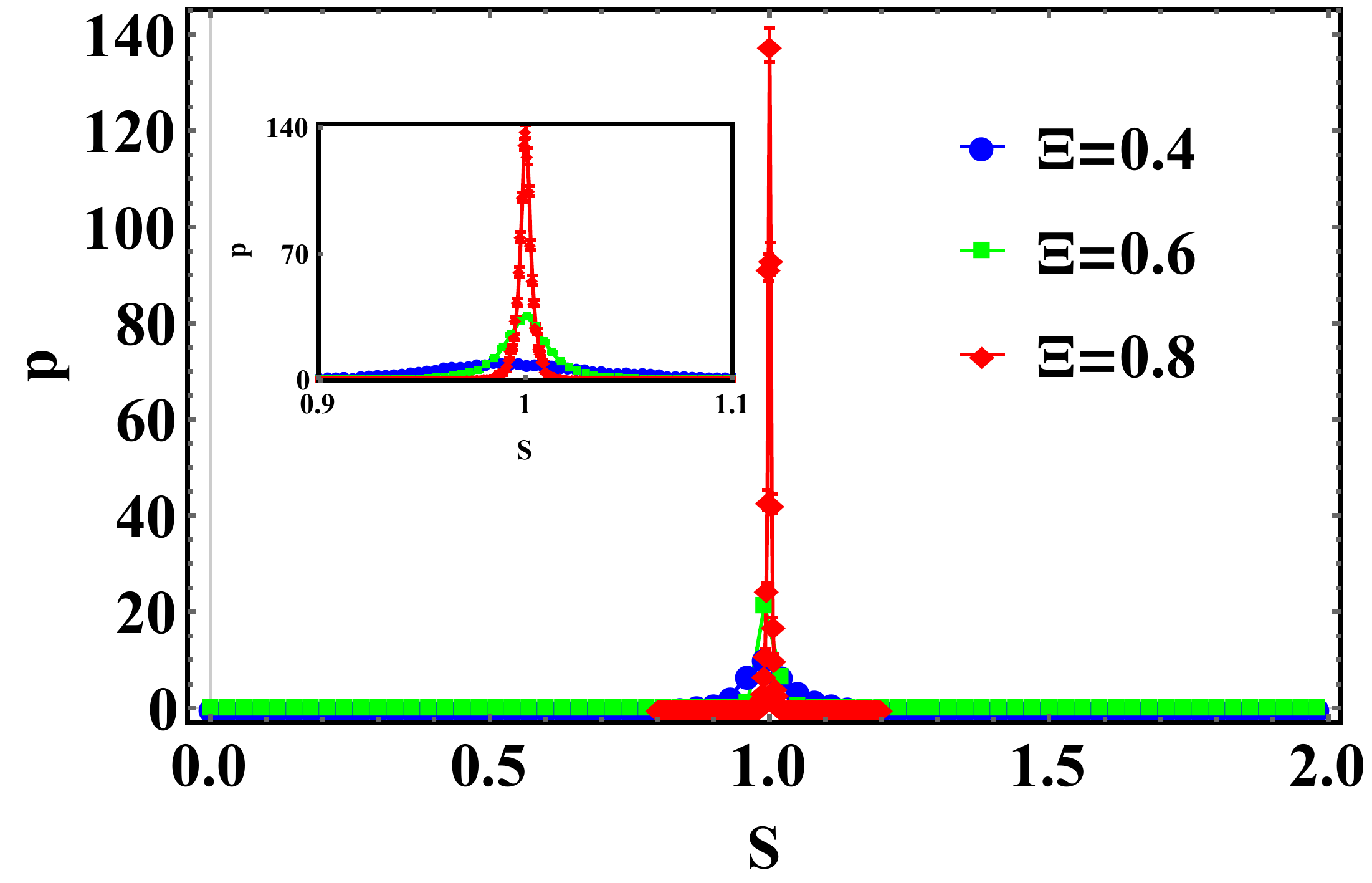}	 
\caption{$\delta$-type level spacing distribution $p(s)$ for the noninteracting 
helical edge model, obtained
by numerical diagonalization of the Hamiltonian Eq.~(\ref{ham-2}) in momentum space. 
We sample every other level to account for the Kramers degeneracy.
Here $s = |\vep_n - \vep_{n-2}|/\Delta$ is the normalized level spacing near energy $\vep_n$, 
while $\Delta$ is the average of $|\vep_n - \vep_{n-2}|$ over the chosen set of levels.
The parameter $\Xi$ 
is the disorder correlation length in units of the inverse ultraviolet momentum 
cutoff \cite{Chou14,SupMat}. 
Inset: The broadening of the $\delta$-type distribution with different values of $\Xi$ \cite{SupMat}.}  
\label{delta-stat}
\end{figure}

\emph{Noninteracting energy level statistics, comparison to Dyson.}---Single particle energy levels
in random systems typically exhibit Poissonian or Wigner-Dyson energy level statistics,
associated to localized or ergodic wave functions \cite{Kravtsov09}.  
Since the 
helical edge
with random RSOC is solved by an integrable transfer matrix, 
we do not expect Wigner-Dyson statistics. 
Using the Heisenberg 
equation of motion for $\Phi(t,x)$ in 
Eqs.~(\ref{tr-phi}) and (\ref{ham-Phi}) with $\td{U} = 0$
gives the static Schr\"odinger equation 
$
	\| \hat{j}(x) \| \hat{J}(-L/2) 
	(-i\d_x)
	\Phi(x) 
	= 
	\vep \Phi(x).
$ 
On a periodic ring with circumference $L$, the eigenenergies 
$\{\vep_n\}$ with $ n \in \mathbb{Z}$ (doubly degenerate) can be obtained via the Bohr-Sommerfeld quantization, 
which leads to $\vep_n = \pm 2 \pi n \Delta_L$. Here the level spacing is
$ \Delta_L^{-1} =\oint_L d x \|\hat{j}(x) \|^{-1}$. For a fixed realization of disorder, 
the energy levels are equally spaced, as for a clean system. 
We confirm this by numerically diagonalizing the original Hamiltonian (\ref{ham-2}) 
in momentum space \cite{Chou14,SupMat}.  
The result is the sharp delta-function-like level-spacing distribution
shown in Fig.~\ref{delta-stat}.

We conclude that the helical edge with a random backscattering 
kinetic
operator 
(induced by RSOC) possesses extended states, shows perfect ballistic transport, and exhibits
clean level statistics. It is interesting to contrast these results for Eq.~(\ref{ham-s}) to 
a nontopological 1D system that incorporates backscattering, but also possesses extended states. 
This is the random mass (``Dyson,'' class BDI) Dirac model \cite{BaFi1997, dyson1997, miev2008}, 
which has the Hamiltonian
\[
	\hat{h}_{\mathsf{Dyson}} = \vf \hat{\sigma}^3 (-i \partial_x) + m(x) \hat{\sigma}^2.
\]
This is particle-hole symmetric in every realization of disorder, and arises 
as the continuum limit of a 1D lattice model perturbed with weakly random nearest-neighbor hopping. 
In sharp contrast to the HLL with random RSOC, however, the extended states
exist only near zero energy (the localization length diverges at $\vep = 0$). Moreover,
an extended Dyson state is quasilocalized, consisting of a \emph{few} isolated peaks with stretched exponential 
tails, separated by large distances \cite{BaFi1997}. 
Because of this rarefied structure the \emph{typical} Landauer conductance decays as 
$G_{\mathsf{typ}} \sim (e^2 / h) \exp(- 2 \sqrt{2 D L / \pi})$, where $D$ is the variance of the random 
mass and $L$ is the length \cite{dyson1997}. 
Position-space profiles of random-RSOC edge state and Dyson wave functions are shown 
in Fig.~\ref{wfs_comparision}. 
For the helical edge state (a), the wave function is ergodic, with only a modulated profile [Eq.~(\ref{psiProfile})].

\emph{Discussion.}---Although the combination of random RSOC and Luttinger interactions does not affect transport, 
in QSH materials the low-temperature conductance will be affected by other scattering mechanisms. 
Finite-temperature corrections, likely to be power law in temperature, may arise due to 
various irrelevant time-reversal symmetric mechanisms such as inelastic 
umklapp processes~\cite{WBZ2006,glazman2012,berkooz2012, mirlin2014, foster2015}, 
phonon scattering~\cite{tra2012b}, or Kondo impurities~\cite{MLO2009,TFM2011,AAY2013,YWYA2015}. More recently, 
it has been suggested that the corrections due to scattering off charge puddles in the bulk~\cite{VGG2013,VGGG2014} 
give rise to a much weaker temperature dependence and might dominate the low temperature transport in existing materials~\cite{VGGG2014}. 
These studies have been performed in terms of the physical $\Psi(x)$ fermion [Eq.~(\ref{ham-s})]. We suggest that, as long as the random 
RSOC is present, it is necessary to carry out calculations in terms of the rotated $\Phi(x)$ fermions [Eq.~(\ref{tr-phi})]. 
We leave this work to future study.


H.-Y.\ X. and M.\ S.\ F. thank M.~M\"{u}ller for useful discussions and acknowledge the 
hospitality of the Kavli Institute for Theoretical Physics (KITP), where this work was partially done (Report No.~NSF-KITP-16-001). 
H.\ L.\ acknowledges hospitality of the Rice University and the Harvard University. 
H.-Y.\ X., Y.-Z.\ C., and M.\ S.\ F. acknowledge funding from the Welch Foundation under Grant 
No. C-1809 and from an Alfred P. Sloan Research Fellowship (No.~BR2014-035).
This research was supported in part by the National Science Foundation under Grant No.~NSF PHY11-25915.



\newpage \clearpage 

\onecolumngrid

\begin{center}
	{\bf \large
	Topological Protection from Random Rashba Spin-orbit Backscattering: \\
	Ballistic Transport in a Helical Luttinger Liquid
	\vspace{4pt}
	\\
	SUPPLEMENTAL MATERIAL
	}
\end{center}

\section{Bosonization of (A) random Rashba spin-orbit coupling and \newline (B) one-particle Umklapp interaction operators}

As explained in the main text, Eq.~(\ref{density}) implies that a time-reversal invariant 
HLL with Rashba spin-orbit coupling is generically 
perturbed by two ``anomalous'' local operators that would be odd under time-reversal in the usual case of a spinless
Luttinger liquid. 
These are the backscattering kinetic operator $\mathcal{O}_{A}(x)$ 
(induced by disorder) 
and 
the one-particle umklapp operator $\mathcal{O}_{B}(x)$
(induced by screened Coulomb interactions),
defined as 
\begin{align}
\tag{S1}
	\label{RRSOC}
	\mathcal{O}_A(x)
	\equiv&\,
	-
	i
	(\zeta_1 + i \zeta_2)
	e^{-2 i k_F x} 
	\left[
		R^\dagger \partial_x L
		-
		(\partial_x R^\dagger) L	
	\right]
	+
	\text{H.c.},
	\\
\tag{S2}
	\mathcal{O}_B(x)
	\equiv&\,
	e^{2 i k_F x}
	L^\dagger R R^\dagger (-i \partial_x) R
	+
	e^{-2 i k_F x}
	R^\dagger L L^\dagger (-i \partial_x) L
	+
	\text{H.c.},
\end{align}	
where H.c.\ denotes the Hermitian conjugate. 
In Eq.~(\ref{RRSOC}), $\zeta_{1,2}$ are arbitrary real parameters. 
We employ field theoretic bosonization conventions \cite{Gogolin,ZinnJustin},
such that the polar U(1) density $\hat{n}$ and current $\hat{I}$ can be expressed as 
\begin{align}\label{BosCurrConv}
\tag{S3}
	\{\hat{n},\hat{I}\} 
	= 
	\frac{1}{\sqrt{\pi}}
	\left\{-\frac{1}{\vf} \partial_t, \vf \partial_x\right\} \phi(t,x)
	=
	\frac{1}{\sqrt{\pi}}
	\left\{\partial_x, - \partial_t\right\} \theta(t,x),
\end{align}
where $\phi$ ($\theta$) denotes the polar (axial) field.
In fact, the first equality holds only in the absence of $\mathcal{O}_{A,B}$,
which modify the form of the current in the polar language. 
The second equality in (\ref{BosCurrConv}) is unaffected; this is an advantage
of the axial field formulation. 

Both operators $\mathcal{O}_A$ and $\mathcal{O}_B$ can be viewed as 
linear combinations of 
Virasoro descendants
\cite{berkooz2012} of conventional Dirac mass terms. 
The bosonic representations can be taken as
\begin{align}
\tag{S4}
	\label{RRSOC--bos}
	\mathcal{O}_A(x)
	\Rightarrow&\,
	\kappa_A
	\,
	\frac{d \phi}{d x}
	\left[
	\zeta_1 \cos(\sqrt{4 \pi} \theta + 2 k_F x)
	+
	\zeta_2 \sin(\sqrt{4 \pi} \theta + 2 k_F x)
	\right],
	\\
\tag{S5}
	\label{1PUInt--bos}
	\mathcal{O}_B(x)
	\Rightarrow&\,
	\kappa_B
	\,
	\frac{d^2 \phi}{d x^2}
	\cos(\sqrt{4 \pi} \theta + 2 k_F x).
\end{align}
In these equations, $\kappa_{A,B}$ denote real constants that 
depend upon the ultraviolet scale and cannot be completely determined in continuum bosonization. 
Similar results were obtained in previous works \cite{japp2010,mirlin2014,foster2015}.
In principle we can also consider a sine partner to Eq.~(\ref{1PUInt--bos}),
but this can be eliminated by shifting the $x$-coordinate origin. 

The time-reversal operation is encoded as 
\[
	\mathcal{T}: \;\; \phi \rightarrow -\phi+\pi/\sqrt{4\pi}, \quad \theta \rightarrow \theta - \pi / \sqrt{4 \pi}, \quad i \rightarrow - i.
\]
Both Eqs.~(\ref{RRSOC--bos}) and (\ref{1PUInt--bos}) are invariant. 
If we perturb the clean HLL with both operators, one can integrate-out the polar field $\phi$ to 
obtain the following axial field action:
\begin{align}\label{HLLRashba}
\tag{S6}
	S 
	=&\,
	\int
	d t \,
	d x 
\left\{
\begin{aligned}
	&\,
	\frac{1}{2 \vc K}
	\left[
	\partial_t \theta 
	-
	\lambda
	\partial_x \cos\left(\sqrt{4\pi}\theta + 2 k_F x \right)
	-
	\xi_1(x) 
	\,
	\sin\left(\sqrt{4\pi}\theta + 2 k_F x \right)
	-
	\xi_2(x) 
	\,
	\cos\left(\sqrt{4\pi}\theta + 2 k_F x \right)
	\right]^2
	\\&\,
	-
	\frac{\vc}{2 K}(\partial_x \theta)^2
	+
	A_{0}
	n
	+
	\frac{1}{c}
	A_{1}
	I
\end{aligned}
\right\},
\end{align}
where we can incorporate an external electric field in $A_\mu$. Here 
$\xi_{1,2}(x)$ are the random RSOC disorder potentials, while $\lambda$ is the coupling strength of the 
one-particle umklapp interaction; 
$\vc$ and $K$ denote the charge velocity and the Luttinger parameter in the absence of the perturbations.


\section{Path integral solution: HLL with random Rashba spin-orbit coupling and Luttinger interactions}

Here we derive Eq.~(18) in the main text. 
We begin with a real time, zero-temperature Grassmann path integral for the HLL with random Rashba spin-orbit coupling and
density-density interactions; the Hamiltonian given by Eq.~(15).
\begin{align}\label{S-Psi}
\tag{S7}
	Z = \int d \bar{\Psi} d \Psi \, \exp(i S),\quad
	S = \int d t \, d x \, 
	\left\{
		i \bar{\Psi} \partial_t \Psi 
		- 
		\bar{\Psi}
		\left[
		\hat{j}(x) \left( -i \d_x  - {\textstyle{\frac{1}{c}}} A_1\right) 
		- 
		{\textstyle{\frac{i}{2}}} 		
		\d_x \hat{j}(x) 
		+ 
		V(x) 
		\right] \Psi
		-
		U(x)
		(\bar{\Psi} \Psi)^2
	\right\},
\end{align}
where we include $A_1$ to incorporate an external electric field. 
We consider an edge spanning $|x| \leq L/2$, connected to ideal leads outside.
For $|x| > L/2$, $\hat{j}(x) = \vf \hat{\sigma}^3$ and $U(x) = 0$. 
A straight-forward bosonization of Eq.~(\ref{S-Psi}) would give the non-linear theory in 
Eq.~(\ref{HLLRashba}) with $\lambda = 0$. 
Instead, we make the change of variables in Eq.~(\ref{tr-phi}), exploiting 
Eq.~(\ref{JTTJ}) to simplify the kinetic term. The result is 
\begin{align}\label{S-Trans}
\tag{S8}
	S = \int d t \, d x \, 
	\left\{
		i \bar{\Phi} \partial_t \Phi 
		- 
		\bar{\Phi}
		\vf \hat{\sigma}^3 \left( -i \d_x - {\textstyle{\frac{1}{c}}} A_1\right) 
		\Phi
		+
		\chi(x)
		\left(
		{\textstyle{\frac{1}{2}}}
		\bar{\Phi}
		i 
		\!
		\stackrel{\leftrightarrow}{\partial}_t 
		\!
		\Phi
		\right)
		-
		\tilde{U}(x)
		(\bar{\Phi} \Phi)^2
	\right\},
\end{align}
where
\begin{align}
\tag{S9}
	\td{U}(x) \equiv \vf^2 \, U(x) / \| \hat{j}(x) \|^2,
	\quad
	\chi(x) \equiv \vf / \| \hat{j}(x) \|  - 1.
\end{align}
In Eq.~(\ref{S-Trans}), 
$a(t) \! \stackrel{\leftrightarrow}{\partial}_t  \! b(t) \equiv a(t) \partial_t b(t) - [\partial_t a(t)] b(t)$.
Using field theoretic bosonization rules \cite{Gogolin,ZinnJustin}, Eq.~(\ref{S-Trans}) can be replaced by the
bosonic action
\begin{align}\label{S-Boson}
\tag{S10}
\begin{aligned}
	S
	=&\,
	\int d t \, d x \, 
	\left\{
	\frac{1}{2 \vf}
	\left[1 + \chi(x)\right]
	\left(\partial_t \theta\right)^2
	-
	\frac{\vf}{2}
	\left[1 - \chi(x) + \frac{2}{\pi \vf}\td{U}(x)\right]
	\left(\partial_x \theta\right)^2
	-
	\frac{1}{\sqrt{\pi} c}
	A_1 
	\partial_t \theta
	\right\}
	\\
	\equiv&\,
	\frac{1}{2}
	\int d t \, d x \, 
	\left[
	\frac{1}{\vc(x) K(x)}
	\left(\partial_t \theta\right)^2
	-
	\frac{\vc(x)}{K(x)}
	\left(\partial_x \theta\right)^2
	-
	\frac{1}{\sqrt{\pi} c}
	A_1 
	\partial_t \theta
	\right],
\end{aligned}
\end{align}
which leads to Eq.~(18). 
In Eq.~(\ref{S-Boson}), $\theta(t,x)$ denotes the axial U(1) field.


\section{Momentum space exact diagonalization}

In this section, we describe diagonalization in momentum space for 
the edge state Hamiltonian with random RSOC given by Eq.~(\ref{ham-s}). 
Scalar potential and Fermi velocity fluctuations are ignored for simplicity. 
The Fourier transform conventions are given by
\begin{align}
\tag{s11}
	\tilde{\psi}_{n}&=\frac{1}{\sqrt{L}}\int dx\, e^{-i\frac{2\pi}{L}nx}\,\psi(x),
	\\
\tag{s12}
	\tilde{\xi}_{a,n}&=\int dx\, e^{-i\frac{2\pi}{L}nx}\,\xi_a(x),
\end{align}
where $n \in \mathbb{Z}$ and $L$ is the system size. We assume periodic boundary conditions.

The Hamiltonian in Fourier space is
\begin{align}
\tag{s13}
	H_0
	=
	\frac{2\pi}{L} v_F 	
	\sum_n\tilde{\psi}^{\dagger}_n\left(n\hat{\sigma}^3\right)\tilde{\psi}_n
	+
	\frac{2\pi}{L^2}
	\sum_{m,n}\tilde{\psi}^{\dagger}_m
	\left[
	\left(\frac{m+n}{2}\right)\left(\xi_{1,m-n}\hat{\sigma}^1+\xi_{2,m-n}\hat{\sigma}^2\right)
	\right]
	\tilde{\psi}_n.
\end{align}

In numerical simulations we have to introduce two scales: 
the cutoff in Fourier modes ($\mathcal{N}$) 
and 
the Gaussian correlation length of the random potentials ($\Xi$). 
The integers $n$ and $m$ are restricted such that $-\mathcal{N}\le n,m\le \mathcal{N}$. 
The total momentum grid has size $(2\mathcal{N}+1)$. 
The momentum cutoff $\Lambda=\frac{2\pi}{L}\mathcal{N}$ is fixed to $2\pi$ in all the simulations. 
We use $r=\frac{L}{\mathcal{N}}=1$ as the finest position-space resolution. 
$v_F$ is also set to one in the simulations.\\

For the random potentials, we assume that
\begin{align}
	\tag{s14}
	\left\langle\xi_1(x)\xi_1(x+R)\right\rangle_{x}=\left\langle\xi_2(x)\xi_2(x+R)\right\rangle_{x}=\Omega_{\xi} \; \kappa(R),
\end{align}
where $\langle\dots\rangle_{x}$ denotes the spatial average, $\kappa(R)$ is a Gaussian distribution function 
(with correlation length $\Xi$), and $\Omega_{\xi}$ is the variance of the random potential. 

\begin{figure}
\centering
\includegraphics[width=0.3\textwidth]{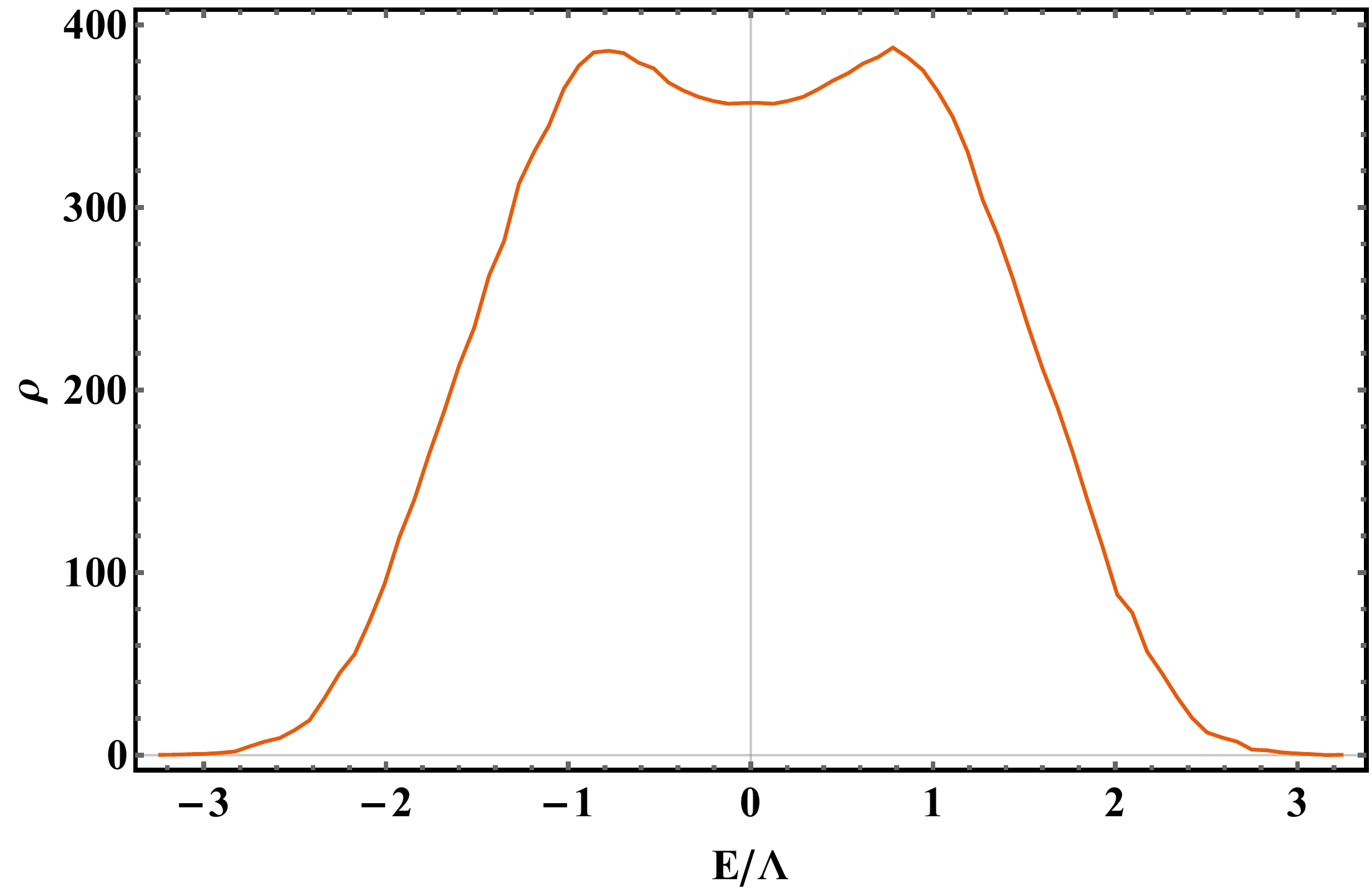}
\includegraphics[width=0.3\textwidth]{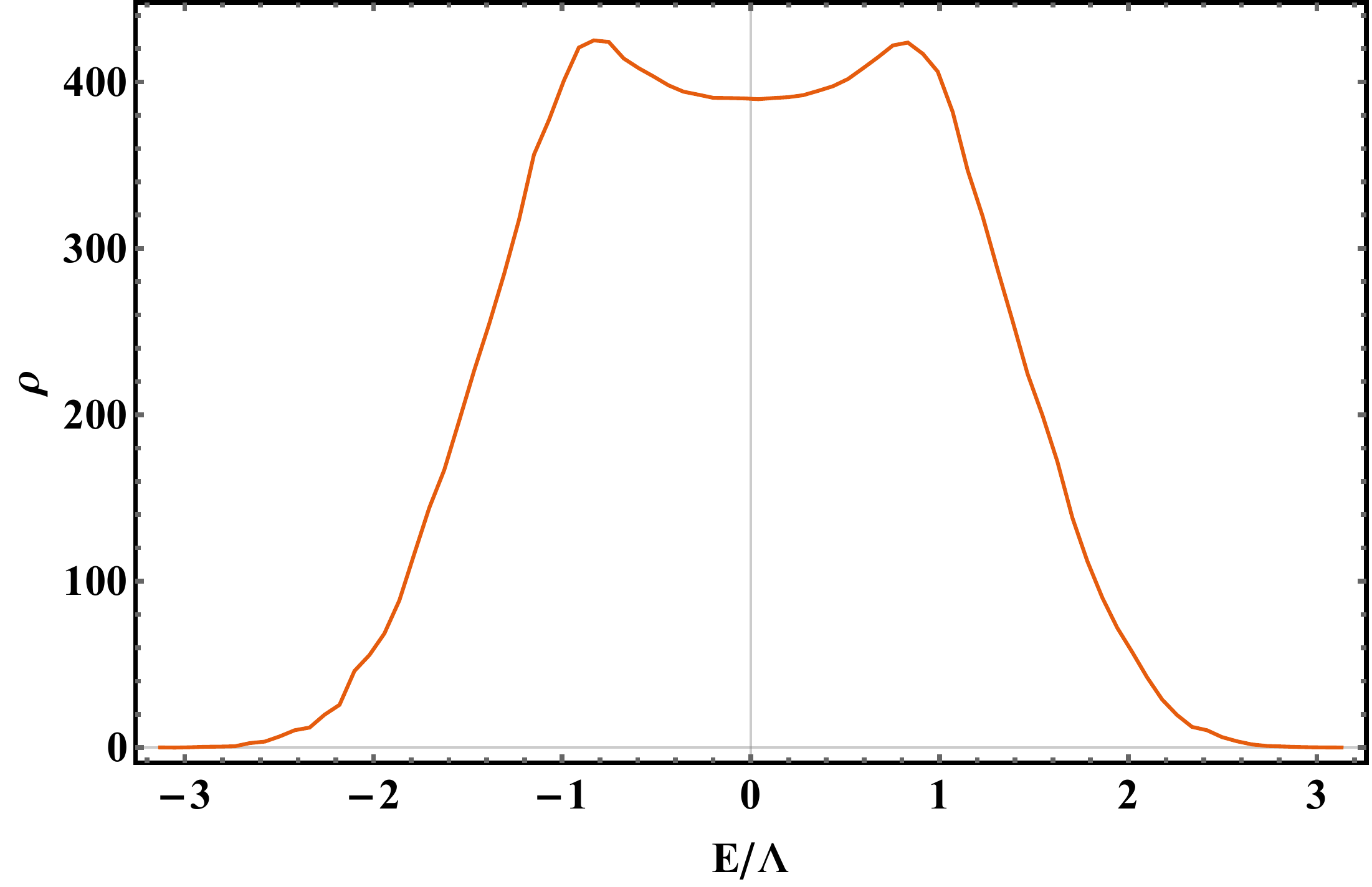}
\includegraphics[width=0.3\textwidth]{dos0p6}
\caption{Density of states of Eq.~(\ref{ham-2}) in the momentum space numerics. 
From left to right: $\Xi=0.4$, $\Xi=0.6$, and $\Xi=0.8$. The variance strength is set to $\Omega_{\xi}=2$ for all the plots. 
The density of states is almost featureless in the interval $-0.5\Lambda\le E\le 0.5\Lambda$. 
The level spacing distributions are constructed from this region.}  
\label{Fig_Sup_DoS}
\end{figure}

We use random phase disorder~\cite{Chou14} to generate the potentials in each fixed realization. 
In Fig.~\ref{delta-stat}, we choose 
$\Omega_{\xi}=2$, 
$L=\mathcal{N}=2000$, 
and 
40 realizations. 
The energy levels are selected from $-0.5\Lambda$ to $0.5\Lambda$ (see Fig.~\ref{Fig_Sup_DoS}) per each realization. 
In Fig.~\ref{delta-stat}, $p(s)$ approaches the delta distribution when the value of 
$\Xi$ is sufficiently large. Our numerical simulation fails for $\Xi$ much smaller than the 
fixed, inverse ultraviolet momentum cutoff 
(white noise limit), due to the truncation of the Fourier spectrum.

\end{document}